\documentstyle[12pt,epsfig]{article}
\begin{document}

\thispagestyle{empty}
\vskip 15pt

\begin{center}
{\Large {\bf Photons from anisotropic {\em Quark-Gluon-Plasma}}}
\renewcommand{\thefootnote}{\alph{footnote}}

\hspace*{\fill}

\hspace*{\fill}

{ \tt{
Lusaka Bhattacharya\footnote{E-mail address:
lusaka.bhattacharya@saha.ac.in}
and
Pradip Roy\footnote{E-mail address:
pradipk.roy@saha.ac.in}
}}\\

\small {\em Saha Institute of Nuclear Physics \\
1/AF Bidhannagar, Kolkata - 700064, INDIA}
\\

\vskip 40pt

{\bf ABSTRACT}
\end{center}

\vskip 0.5cm

We calculate medium photons due to Compton and annihilation processes 
in an anisotropic media. The effects of time-dependent momentum-space 
anisotropy of {\em Quark-Gluon-Plasma} (QGP) on the medium photon 
production are discussed. Such an anisotropy can results from the 
initial rapid longitudinal expansion of the matter, created in relativistic 
heavy ion collisions. A phenomenological model for the time-dependence 
of the parton hard momentum scale, $p_{hard}(\tau)$, and anisotropy 
parameter, $\xi(\tau)$, has been used to describe the plasma space-time 
evolution. We find significant dependency of photon yield 
on the isotropization time ($\tau_{iso}$). It is shown that the 
introduction of early time momentum-space anisotropy can 
enhance the photon production by a factor of $10~(1.5)$ 
(in the central rapidity region) for {\em free streaming} 
({\em collisionally-broadened}) {\em interpolating} model if we assume 
fixed initial condition. On the other hand, enforcing the fixed 
final multiplicity significantly reduces the enhancement of medium 
photon production.   
 
\vskip 30pt

\section{Introduction}

One of the goals for the ongoing relativistic heavy ion 
collision experiments at the Relativistic Heavy Ion 
Collider (RHIC) and the upcoming experiments at CERN Large Hadron 
Collider (LHC) is to produce and study the properties of QGP. 
According to the prediction 
of lattice quantum chromodynamics, QGP is expected to 
be formed when the temperature of nuclear matter is raised above 
its critical value, $T_c\sim 170$ MeV, or equivalently the energy 
density of nuclear matter is raised above $1~GeV/fm^{3}$
~\cite{PRC72_ref1}. The possibility of QGP formation at RHIC 
experiment, with $5~GeV/fm^{3}$ initial density of the created 
system, is supported by the observation of high $p_T$ hadron suppression
(jet-quenching) in the central Au-Au collisions compared to the 
binary-scaled hadron-hadron collisions~\cite{jetquen}. 
Apart form jet-quenching, 
several possible probes have been studied in order to characterize the 
properties of QGP.

However, many properties of QGP are still poorly understood. The most 
debated question is whether the matter formed in the relativistic heavy ion
collisions is in thermal equilibrium or not. The measurement of elliptic
flow parameter and its theoretical explanation suggest that the matter
quickly comes into thermal equilibrium 
(with $\tau_{\rm therm} < $ $1$ fm/c, where $\tau_{therm}$ is the time of 
thermalization)~\cite{PRC75_ref1}.
On the contrary, perturbative estimation suggests the slower thermalization of
QGP~\cite{PRC75_ref2}. However, recent hydrodynamical 
studies~\cite{0805.4552_ref4} have shown that 
due to the poor knowledge of the initial conditions there is a
sizable amount of uncertainty in the estimate of thermalization or 
isotropization time. It is suggested that (momentum) anisotropy driven
plasma instabilities may speed up the process of 
isotropization~\cite{plb1181993}, in that case one is allowed to use 
hydrodynamics for the evolution of the matter. However, instability-driven 
isotropization is not yet proved at RHIC and LHC energies. 

In view of the absence of a theoretical proof behind the rapid
thermalization and the uncertainties in the hydrodynamical fits of
experimental data, it is very hard to assume hydrodynamical
behavior of the system from the very beginning. Therefore, one is
forced to find out some observables which are sensitive to the
early time after the collision. Electromagnetic probes have long
been considered as the most effective one to characterize
the initial stages of heavy ion collisions. 
Photons can be one of such observables. Since photons interact only 
electromagnetically with the plasma, they remains relatively less 
affected by the later stages of the plasma evolution. Therefore, 
photons can carry information about the 
plasma initial conditions~\cite{janejpg,dksepjc,pasi} only if the observed
flow effects from the late stages of the collisions can be understood
and modeled properly. 
The observation of pronounced transverse flow in the photon transverse
momentum distribution has been taken into account in model calculations of
 photon $p_T$ distribution
at various beam 
energies~\cite{janejpg,dksepjc,clem,janerap,trenk-hep-ph/0408218}. 
It is found that because
of the transverse kick the low energy photons populate the intermediate
regime and consequently, the contribution from hadronic matter becomes
comparable with that from the hadronic matter destroying the window
where the contribution from QGP was supposed to
dominate~\cite{trenk-hep-ph/0408218}.
Apart from transverse flow effects,
the investigation of longitudinal evolution using HBT correlation
measurements has been done in Ref.~\cite{prc71064905}. It is shown that the
decrease of $R_{\rm side}$ with $p_T$ ($>$ 2.5 GeV) provides a good
indication whether transverse flow is significant or not. However,
so far as the pre-equilibrium emission is concerned this effect is
not important.
Moreover,
the elliptical flow ($v_2^{\gamma}$) of photons has recently been calculated in
Refs.~\cite{npa783379c2007,turbideprl} where it is shown that $v_2^{\gamma}$
has a rich structure which reflect the interplay between different emission
processes. It is argued that the total photon elliptical flow is
small at high $p_T$ implying small elliptical flow during the early stages
of the collisions. 

For the above mentioned reason photon production from relativistic 
heavy ion collisions assuming isotropic QGP has been extensively 
studied~\cite{kapusta,jane}. 
In all these works it is assumed that the plasma thermalizes rapidly with 
$\tau_{therm}= \tau_{i}$ where $\tau_{i}$ is the time 
scale of parton formation. In spite of equating thermalization time 
to the parton formation time, in this work we will
introduce an intermediate time scale (isotropization time,
$\tau_{iso}$) to study the effects of early time momentum-space anisotropy 
on the medium photon production. The parton formation time $\tau_i$ can be 
estimated from the nuclear saturation scale, i. e. 
$\tau_i \sim Q_s^{-1}$~\cite{raju}, where $Q_s \sim 1.5~(2) $ GeV 
for RHIC (LHC). 
Recently, it has been shown in 
Ref.~\cite{mauricio} that for fixed initial conditions, the introduction 
of a pre-equilibrium momentum-space anisotropy enhances high energy 
dileptons by an order of magnitude. To include the time dependent 
momentum-space anisotropy, we will use the phenomenological model 
of Ref.~\cite{mauricio,prl100}. This model assumes two time 
scales: the parton formation time, $\tau_i$, and the isotropization time,
$\tau_{iso}$, which is the time when the system becomes isotropic 
in momentum space. Immediately after the formation of QGP the system
can be assumed to be isotropic~\cite{5of4552}. Subsequent rapid
expansion of the matter along the beam direction causes faster cooling in
the longitudinal direction than in the transverse
direction~\cite{PRC75_ref2}. As a result, the system becomes anisotropic
with $\langle{p_L}^2\rangle << \langle{p_T}^2\rangle$ in the local rest frame. 
At some later time when the effect of parton interaction rate 
overcomes the plasma expansion rate, the system returns to the 
isotropic state again and remains isotropic for the rest of the period.


The plan of the paper is the following. We will discuss the medium 
photon production rate and the space-time evolution of anisotropic 
QGP in the next section. Section 3 is devoted  to describe the results for
various evolution scenarios. We summarize in section 4.

\section{Formalism}
\subsection{Photon rate} 
\begin{figure}[t]
\begin{center}
\epsfig{file=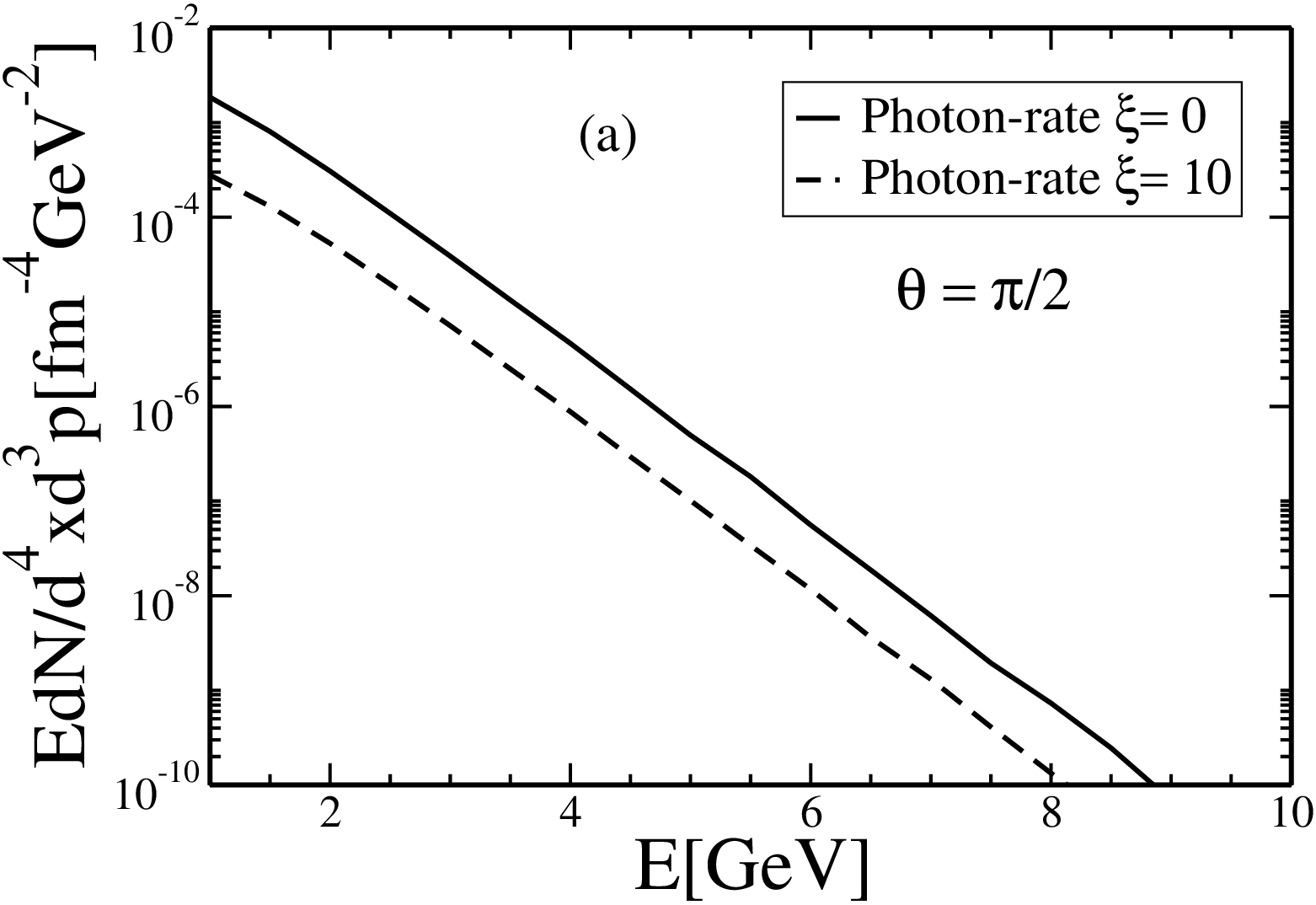,width=7cm,height=8cm,angle=270}
\epsfig{file=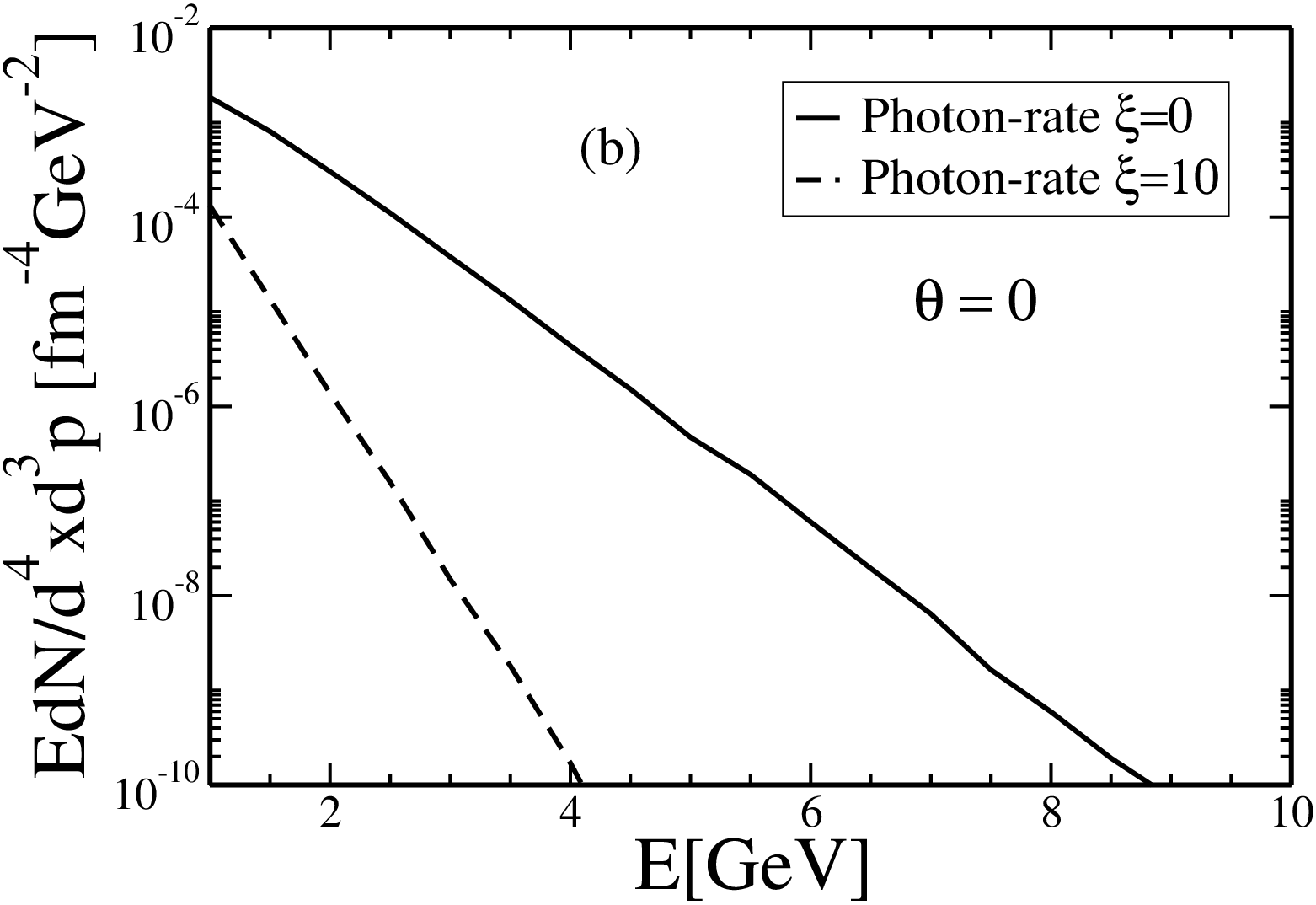,width=7cm,height=8cm,angle=270}
\end{center}
\caption{Photon rates for photon propagation angle 
(a) $\theta =\pi/2$ and (b) $\theta =0$ as a function of
  photon energy ($E$) for two different value of anisotropy
  parameter $\xi$ with $\alpha_{s}=0.3$ and $p_{hard}=0.446$ GeV}
\label{fig1}
\end{figure}
\begin{figure}[h]
\begin{center}
\epsfig{file=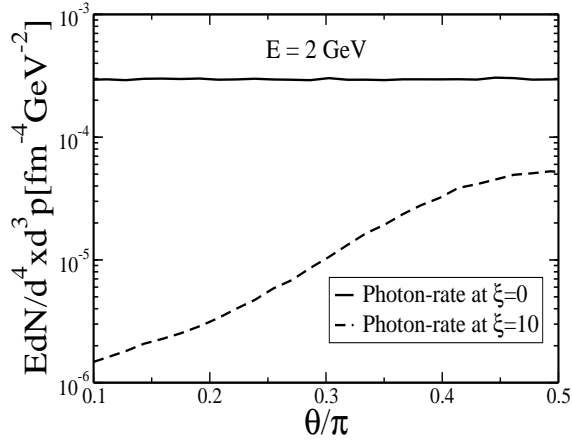,width=7cm,height=8cm,angle=270}
\end{center}
\caption{Photon rate as a function of
  photon propagation angle ($\theta$) for two different 
value of anisotropy parameter $\xi$ with $\alpha_{s}=0.3$ 
and $p_{hard}=0.446$ GeV for $E=2$ GeV.}
\label{fig2}
\end{figure}
\begin{figure}[h]
\begin{center}
\epsfig{file=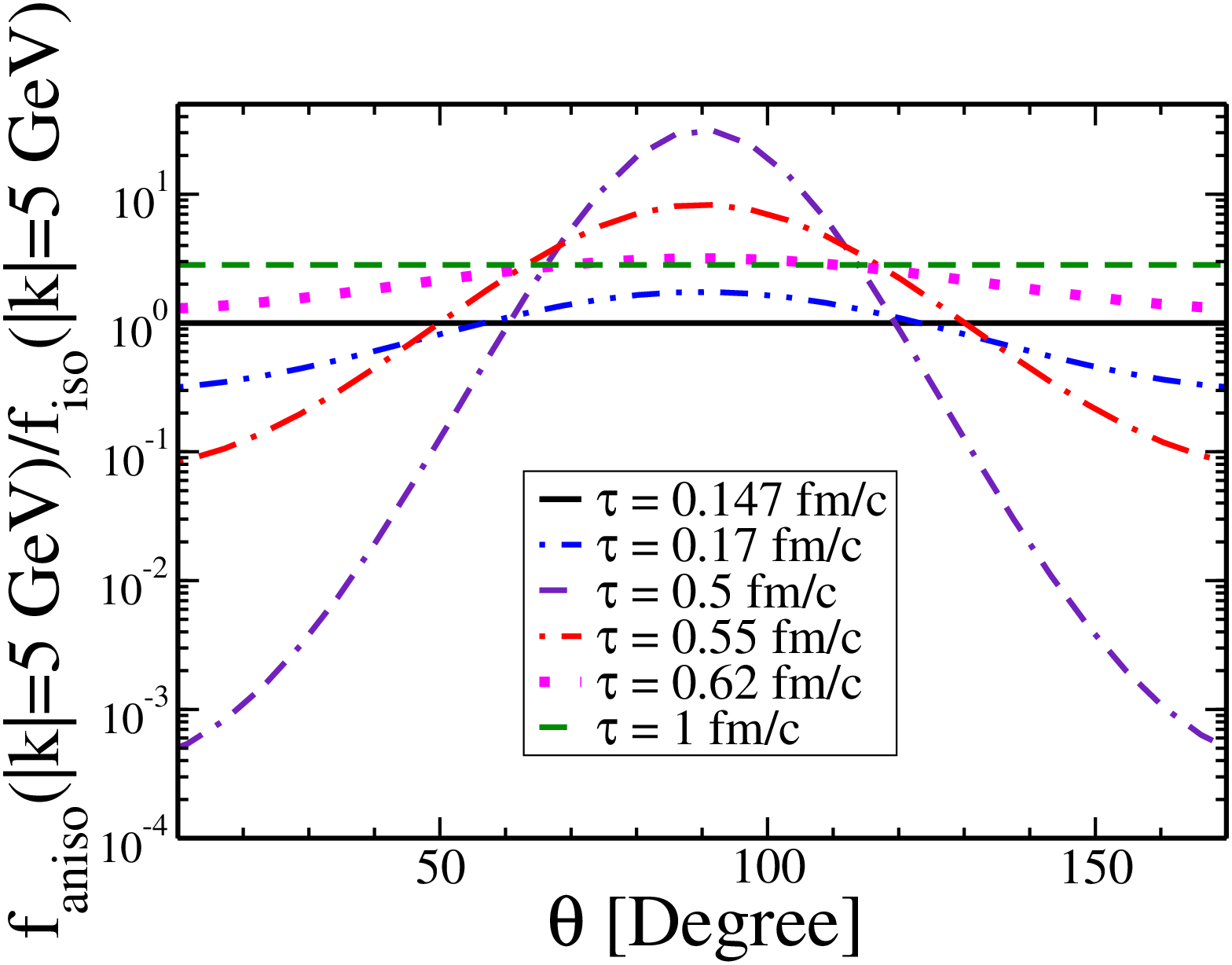,width=7cm,height=8cm,angle=270}
\end{center}
\caption{(Color online) 
${f_{aniso}(|{\bf k}|=5~GeV)}/{f_{iso}(|{\bf k}|=5~GeV)}$ 
as a function of angle ($\theta$) with direction of
  anisotropy for $\tau_{iso}=0.5$ fm. Different curves correspond to 
the ratio at different intermediate time.}
\label{fig3}
\end{figure}

The lowest order processes for photon emission from QGP are the
Compton ($q ({\bar q})\,g\,\rightarrow\,q ({\bar q})\,
\gamma$) and the annihilation ($q\,{\bar q}\,\rightarrow\,g\,\gamma$)
processes. The total cross-section (for both the processes) diverges in the 
limit $t/u\to 0$. These singularities have to be shielded by thermal
effects in order to obtain infrared safe calculations. It has been argued
in Ref.~\cite{kajruus} that the intermediate quark  acquires a thermal
mass in the medium, whereas the hard thermal loop (HTL) approach of
Ref.~\cite{Brapi} shows that very soft modes are suppressed in a
medium providing the natural cut-off $k_c \sim {\sqrt g}T$.

The rate of photon production ($EdN/d^4xd^3p$) from 
anisotropic plasma due to Compton and annihilation processes has been 
calculated in Ref.~\cite{prd762007}. The soft contribution is calculated by
evaluating the photon polarization tensor for an oblate momentum-space
anisotropy of the system where the cut-off scale is fixed at 
$k_c \sim  \sqrt g p_{hard}$. 
Here $p_{hard}$ is a hard-momentum scale that appears in the distribution
functions. 

In this work we assume that the infrared singularities can be shielded by the
introduction of the thermal masses for the participating partons.
This is a good approximation at times short compared to the time scale
when plasma instabilities start to play an important role. We shall see
that as long as the static rate is concerned our results are similar
to that of Ref.~\cite{prd762007}.
The differential photon production rate for $1+2\to3+\gamma$ processes in 
an anisotropic medium is given by~\cite{prd762007} 
\begin{eqnarray} 
E\frac{dN}{d^4xd^3p}&=& 
\frac{{\mathcal{N}}}{2(2\pi)^3} 
\int \frac{d^3p_1}{2E_1(2\pi)^3}\frac{d^3p_2}{2E_2(2\pi)^3}
\frac{d^3p_3}{2E_3(2\pi)^3}
f_1({\bf{p_1}},p_{\rm hard},\xi)f_2({\bf{p_2}},p_{\rm hard},\xi) \nonumber\\
&\times&(2\pi)^4\delta(p_1+p_2-p_3-p)|{\mathcal{M}}|^2 
[1\pm f_3({\bf{p_3}},p_{\rm hard},\xi)]
\label{photonrate}
\end{eqnarray}
where, $|{\mathcal{M}}|^2$ represent the spin averaged matrix element
squared for one of those processes which contributes in the photon rate
and ${{\mathcal N}}$ is the degeneracy factor of the corresponding
process. $\xi$ is a parameter controlling the strength of the anisotropy 
with $\xi > -1$. $f_1$, $f_2$ and $f_3$ are the anisotropic 
distribution functions of the medium partons and will be 
discussed in the following.   

In an isotropic system particles move in all directions with equal
probability. However, for an anisotropic system there will be atleast
one preferred direction. The anisotropic distribution function can be
obtained~\cite{stricland} by squeezing or stretching an arbitrary
isotropic distribution function along the preferred direction in 
momentum space,
\begin{eqnarray}
f_{i}({\bf k},\xi, p_{hard})=f_{i}^{iso}(\sqrt{{\bf k}^{2}+\xi 
({\bf k.n})^{2}},p_{hard})
\label{dist_an}
\end{eqnarray}
where ${\bf n}$ is the direction of anisotropy. It is important to
notice that $\xi > 0$ corresponds to 
a contraction of the distribution function in the direction of 
anisotropy and $-1 < \xi < 0 $ corresponds to a stretching in the
direction of anisotropy. In the context of relativistic
heavy ion collisions, one can identify the direction of anisotropy with 
the beam axis along which the system expands initially. The hard momentum
scale $p_{hard}$ is directly related to the average momentum of the 
partons. In the case of an isotropic QGP,
$p_{hard}$ can be identified with the plasma temperature ($T$).

\subsection{Space time evolution}

\begin{figure}[h]
\begin{center}
\epsfig{file=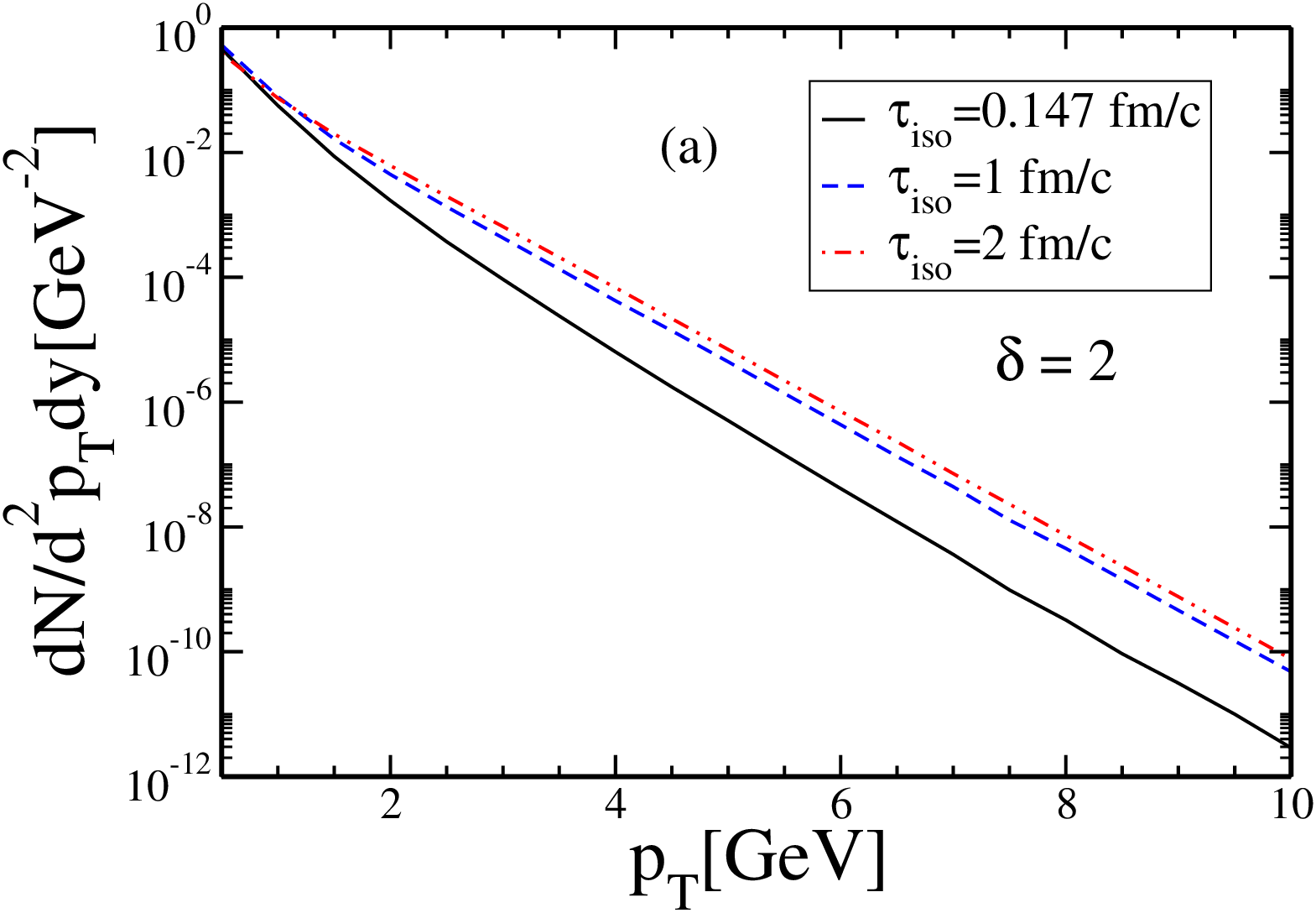,width=7cm,height=8cm,angle=270}
\epsfig{file=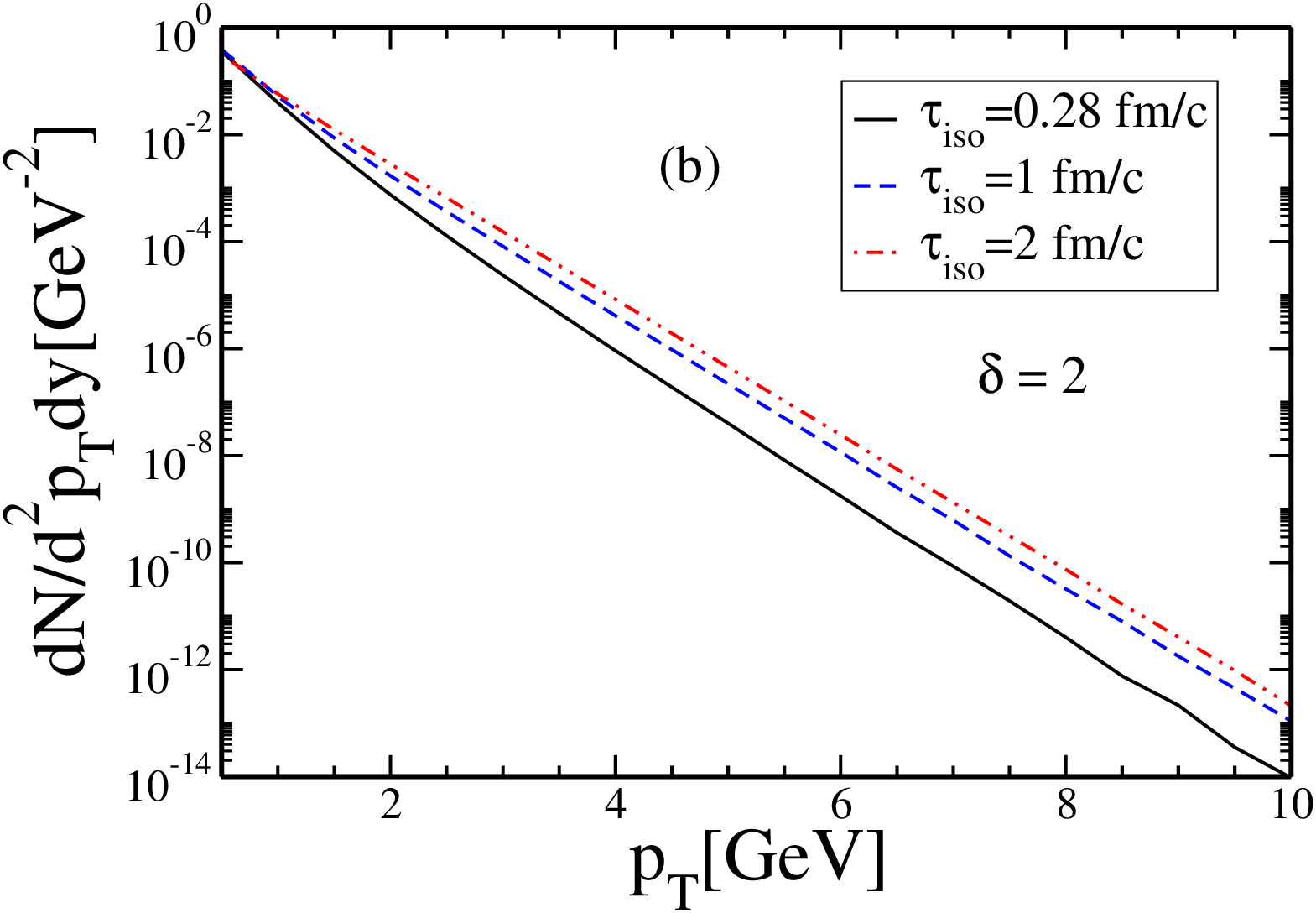,width=7cm,height=8cm,angle=270}
\end{center}
\caption{(Color online) Medium photon spectrum, $dN/dyd^2p_T$, at
$\theta=\pi/2$ ($y=0$) for the {\em free-streaming interpolating }
model ($\delta = 2$) for three different values of
isotropization time, $\tau_{iso}$, with initial conditions, 
(a) Set-I  and (b) Set-II .}
\label{fig4}
\end{figure}
\begin{figure}[h]
\begin{center}
\epsfig{file=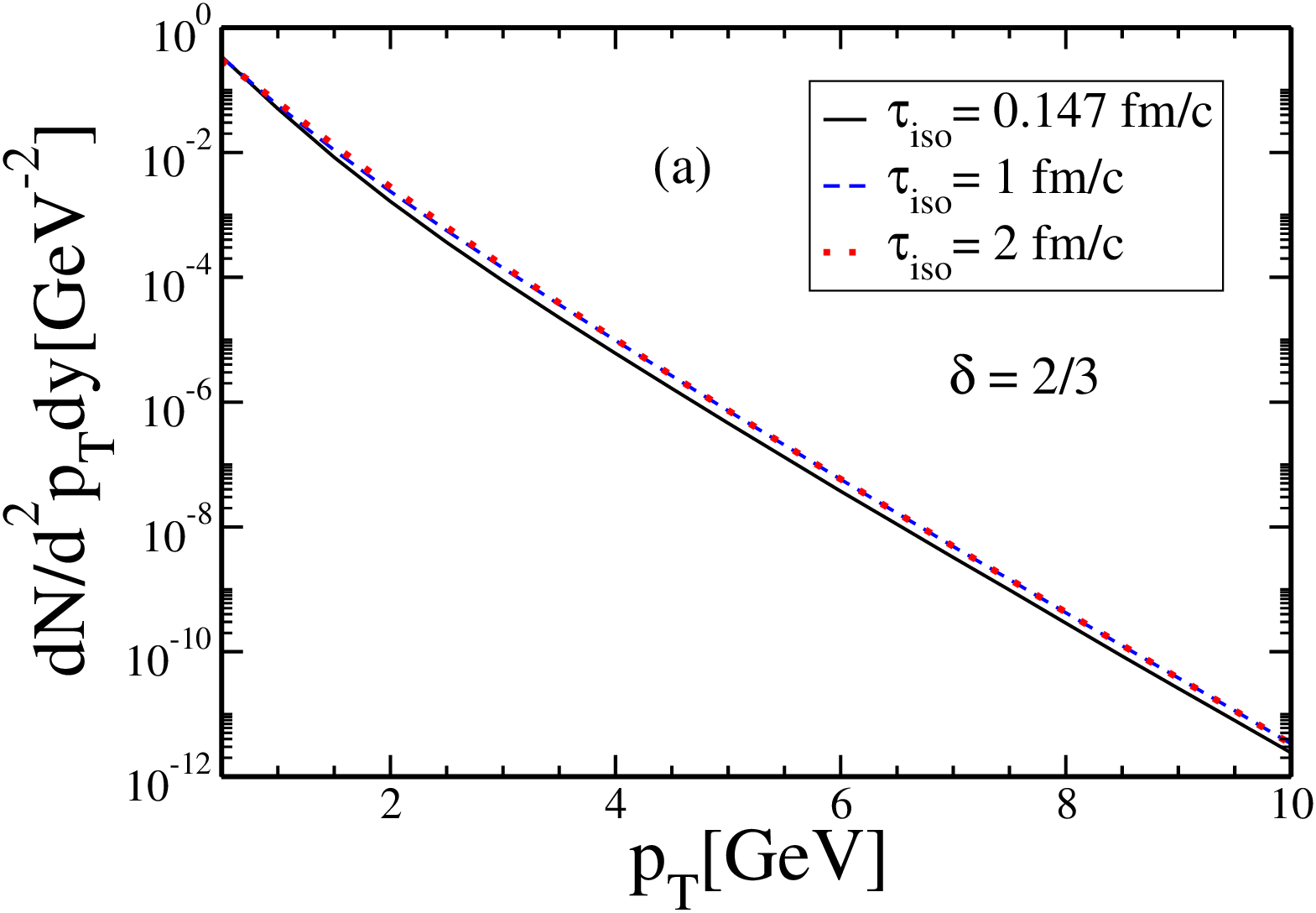,width=7cm,height=8cm,angle=270}
\epsfig{file=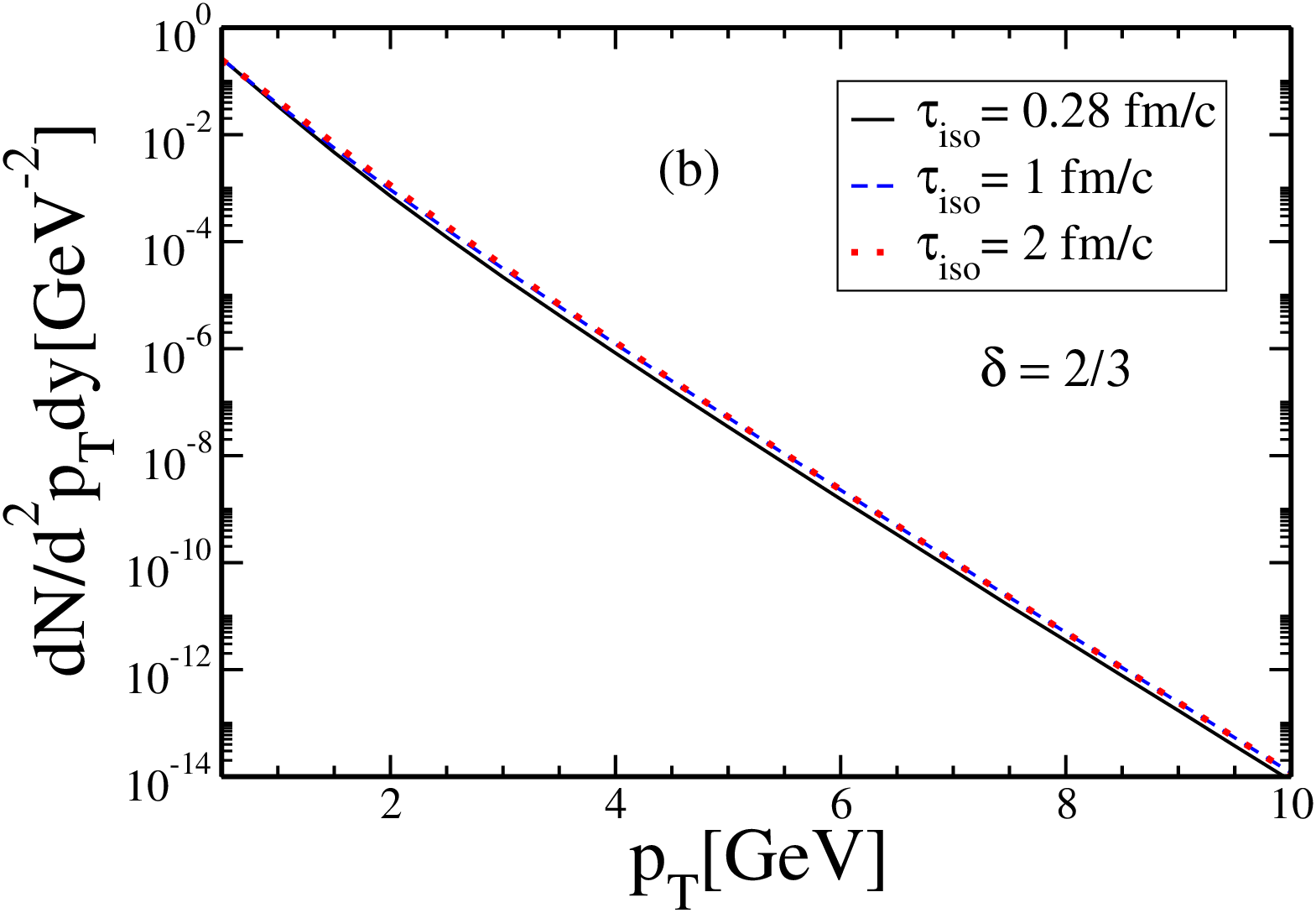,width=7cm,height=8cm,angle=270}
\end{center}
\caption{(Color online) Same as Fig.~\protect\ref{fig4} with $\delta=2/3$.}  
\label{fig5}
\end{figure} 
\begin{figure}[t]
\begin{center}
\epsfig{file=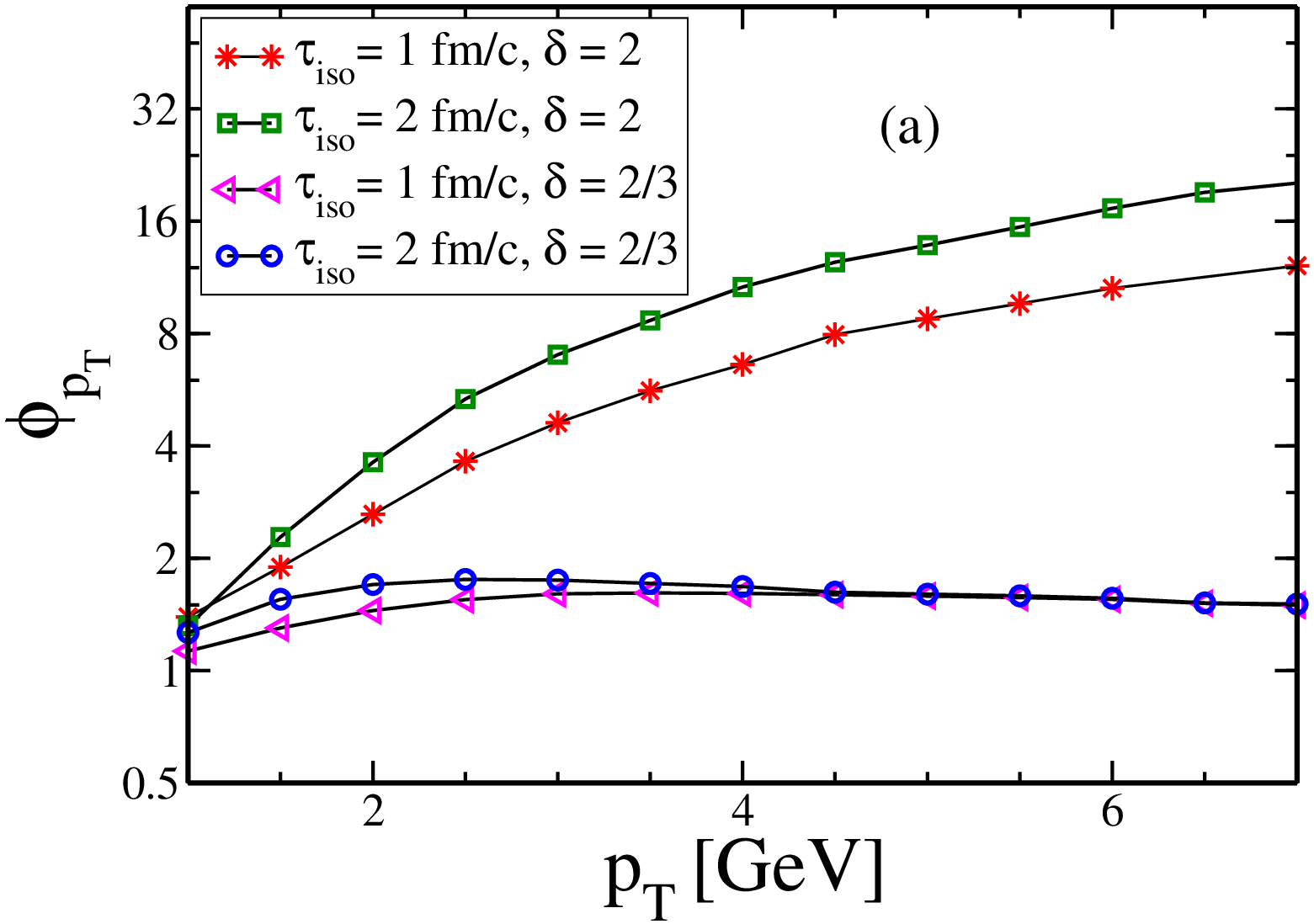,width=7cm,height=8cm,angle=270}
\epsfig{file=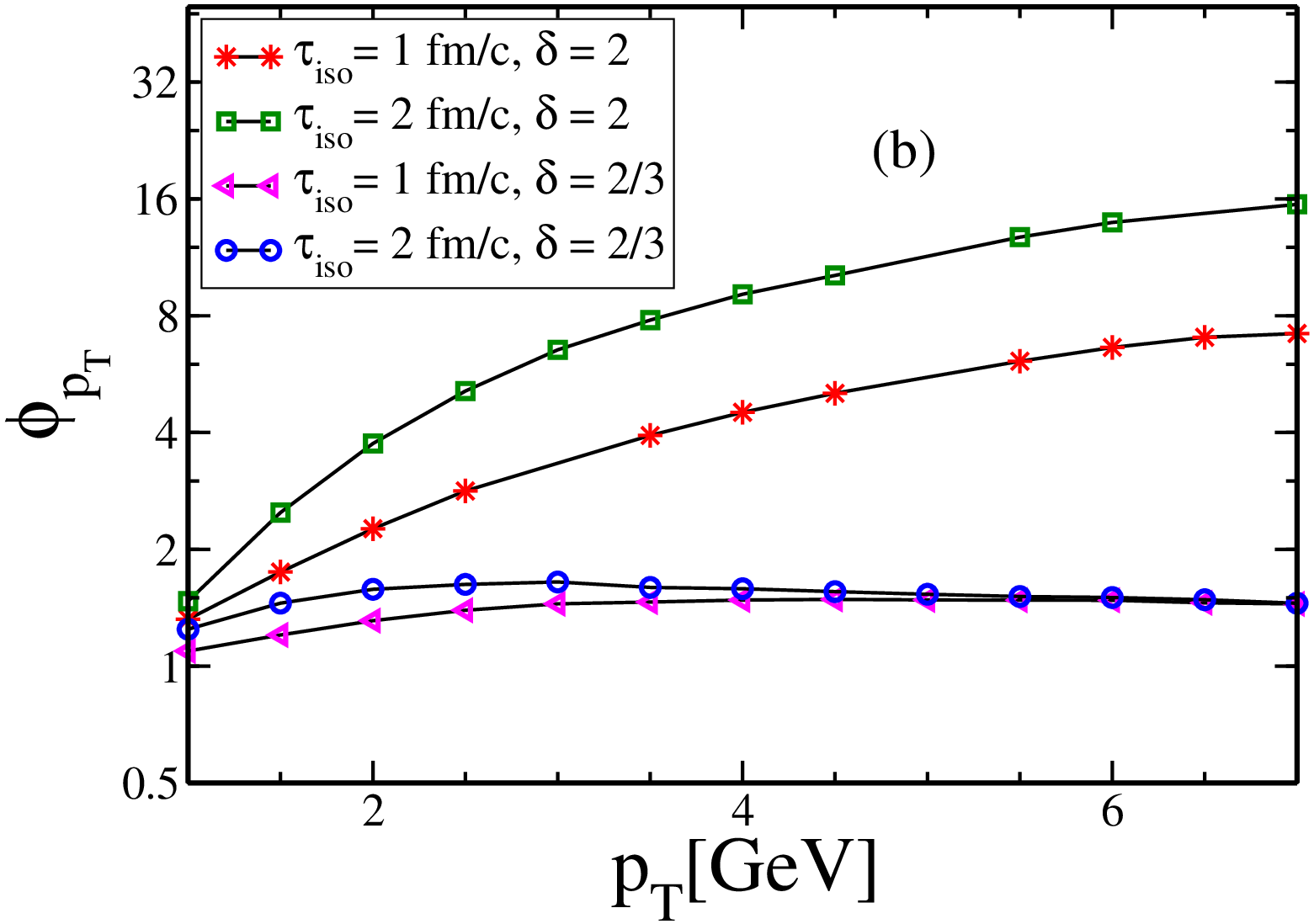,width=7cm,height=8cm,angle=270}
\end{center} 
\caption{{(Color online) \em Photon enhancement factor} 
$(\phi_{p_T})$ as a function of photon transverse momentum $(p_T)$ for the 
{\em free-streaming} and {\em collisionally-broadened interpolating} models 
with fixed initial conditions. (a) and (b) correspond to Set-I and 
Set-II respectively.}  
\label{fig6}
\end{figure} 

The previous discussion on the medium photon rate is not enough to make
any phenomenological prediction for the expected medium photon yield
in an anisotropic QGP as it gives the static photon rate which is also
calculated in Ref.~\cite{prd762007}. In other words, the rate is not
convoluted with the evolution of the matter. To do this
one needs to know the time dependence of $p_{\rm hard}$ and $\xi$. 
We shall follow the work of Ref.~\cite{mauricio} to evaluate the
$p_T$ distribution of photons from the first few Fermi of the 
plasma evolution. Three scenarios of the space-time evolution (as 
described in Ref.~\cite{mauricio}) are the following:
(i) $\tau_{\rm iso} = \tau_i$, the system evolves hydrodynamically
so that $\xi =0$ and $p_{\rm hard}$ can be identified with the
temperature ($T$) of the system (till date all the calculations have been
performed in this scenario), (ii) $\tau_{\rm iso}\rightarrow \infty$,
the system never comes to equilibrium, 
(iii) $\tau_{\rm iso} > \tau_i$ and $\tau_{\rm iso}$ is finite, one should
devise a time evolution model for $\xi $ and $p_{\rm hard}$ which smoothly 
interpolates between pre-equilibrium anisotropy and hydrodynamics. We 
shall follow scenario (iii) (see Ref.~\cite{mauricio} for details) 
in which case the time dependence of the anisotropy parameter $\xi$ is 
given by
\begin{eqnarray}
\xi(\tau,\delta) &=& (\frac{\tau}{\tau_i})^\delta-1 
\label{eq_xi}
\end{eqnarray}
where the exponent $\delta = 2~(2/3)$ corresponds to {\em free-streaming 
(collisionally-broadened)} pre-equilibrium momentum space anisotropy and
$\delta=0$ corresponds to thermal equilibrium. 
As in Ref.~\cite{mauricio}, a transition width $\gamma^{-1}$ is introduced 
to take into account the smooth transition from non-zero value of $\delta$ 
to $\delta = 0$ at $\tau = \tau_{\rm iso}$. The time dependence of 
various quantities are, therefore, obtained in terms of a smeared step 
function \cite{prl100}:
\begin{equation} 
\lambda(\tau)=\frac{1}{2}(\tanh[\gamma(\tau-\tau_{iso})/\tau_i]+1). 
\label{eq_lamda}
\end{equation}
For $\tau << \tau_{\rm iso} ( >> \tau_{\rm iso})$ we have $\lambda = 0 (1)$
which corresponds to {\em free streaming} (hydrodynamics). With this, the 
time dependence of relevant quantities are as follows~\cite{mauricio}:
\begin{eqnarray}
\xi(\tau,\delta) &=& \left(\frac{\tau}{\tau_i}\right)^
{\delta(1-\lambda(\tau))}-1,\nonumber\\
p_{\rm hard}(\tau)&=&T_i~{\bar {\cal U}}^{1/3}(\tau),
\label{xirho}
\end{eqnarray}
where, 
\begin{eqnarray}
{\mathcal U}(\tau)&\equiv& \left[{\mathcal
    R}\left((\frac{\tau_{\rm iso}}{\tau})^\delta-1\right)
\right]^{3\lambda(\tau)/4}
\left(\frac{\tau_{\rm iso}}{\tau}\right)^{1-\delta(1-\lambda(\tau))/2},
\nonumber\\
{\bar {\cal U}}&\equiv& \frac{{\cal U}(\tau)}{{\bar {\cal U}}(\tau_i)},
\nonumber\\
{\mathcal R}(x)&=&\frac{1}{2}[1/(x+1)+\tan^{-1}{\sqrt
    {x}}/\sqrt{x}]
\label{utau}
\end{eqnarray}
and $T_i$ is the initial temperature of the plasma. In our 
calculation, we assume a fast-order phase transition beginning at the 
time $\tau_f$ and ending at $\tau_{H}=r_d\tau_f$ 
where $r_d=g_Q/g_H$ is the ratio of the degrees of freedom in the two 
(QGP phase and hadronic phase) phases and $\tau_f$ is obtained by the 
condition $p_{\rm hard}(\tau_f) = T_c$ which we take as $170$ MeV. 
We also include the contribution from the mixed phase. 
Therefore, the total medium photon yield, arising from pure QGP phase 
and mixed phase is given by, 
\begin{equation}
\frac{dN^{\gamma}}{dyd^2p_T}=\pi
{R_{\perp}}^2~\left[\int_{\tau_i}^{\tau_f}\tau d\tau~\int d\eta~
\frac{dN^{\gamma}}{d^4xdyd^2p_T} + \int_{\tau_f}^{\tau_H}f_{QGP}(\tau)\tau
d\tau~\int d\eta~\frac{dN^{\gamma}}{d^4xdyd^2p_T}\right],  
\label{yield_total} 
\end{equation}
where, $f_{QGP}(\tau)=(r_d-1)^{-1}(r_d \tau_f \tau^{-1}-1)$, is the
fraction of the QGP phase in the mixed phase~\cite{PRC72_38}, 
$R_{\perp}=1.2A^{1/3}$ fm is the radius of the colliding 
nucleus in the transverse plane. The energy of the 
photon in the fluid rest frame is given by 
$E_{\gamma}=p_T cosh(y-\eta)$ where $\eta$ and $y$ are the 
space-time and photon rapidities respectively.

\section{Result}
\begin{figure}[h]
\begin{center}
\epsfig{file=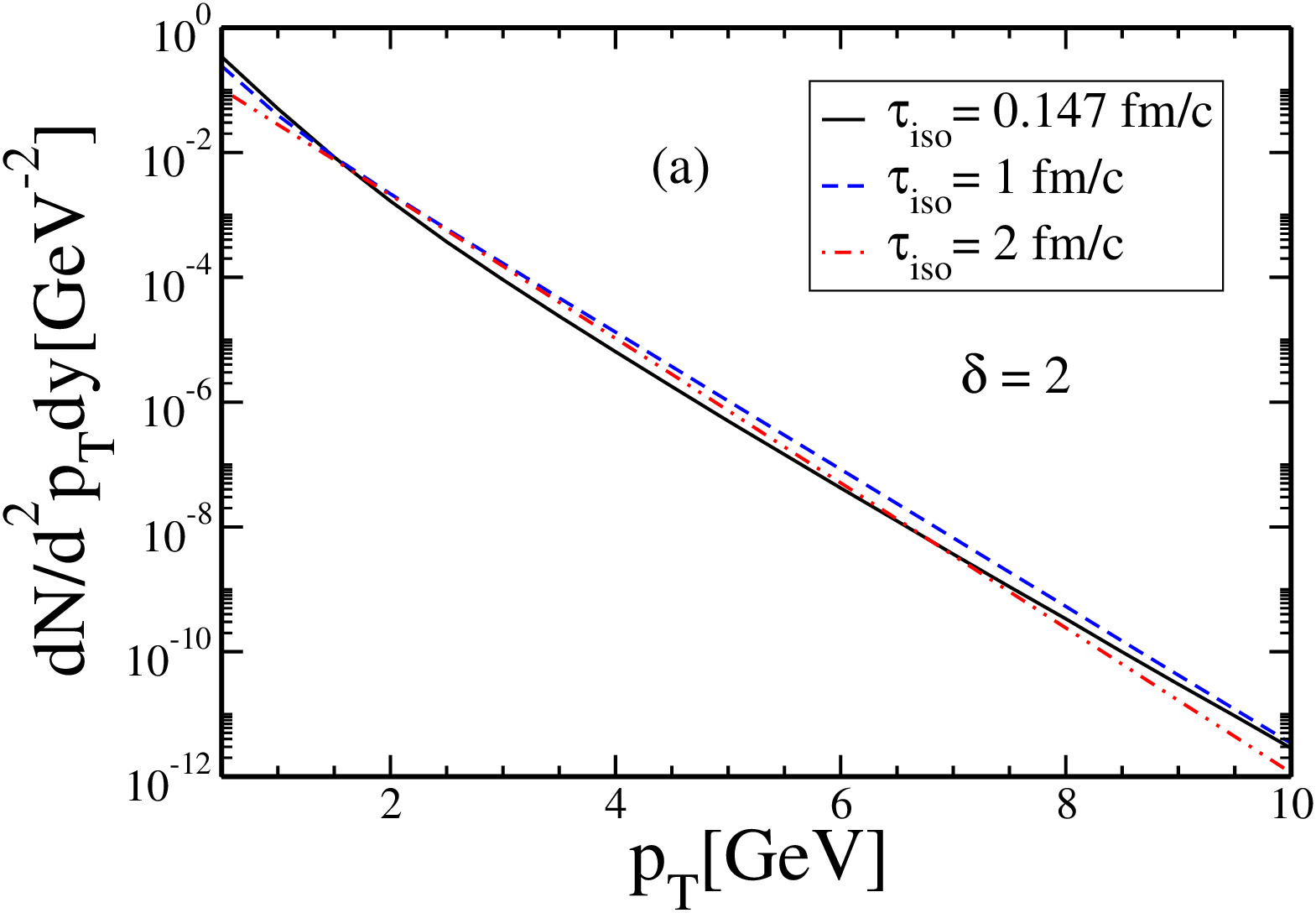,width=7cm,height=8cm,angle=270}
\epsfig{file=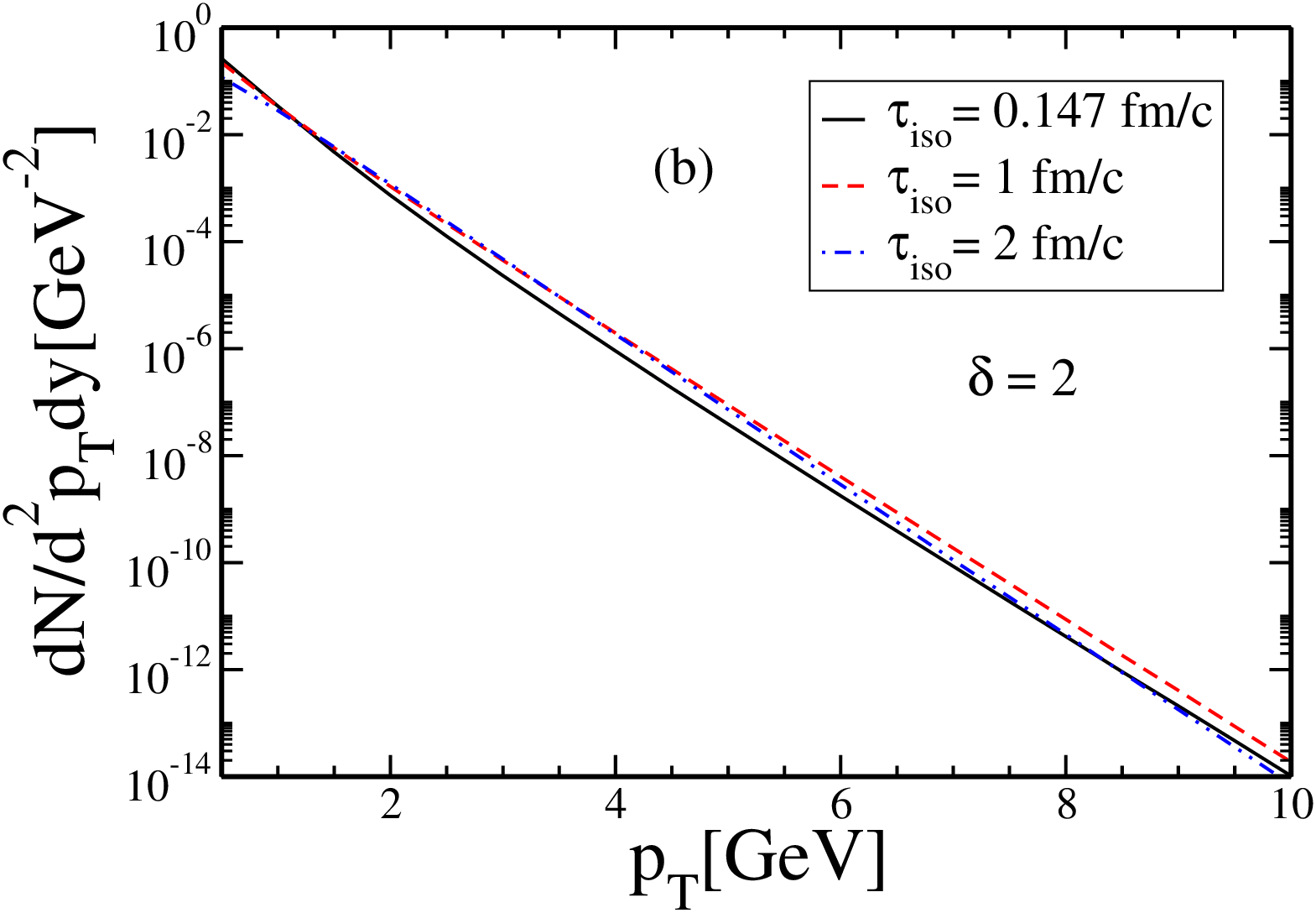,width=7cm,height=8cm,angle=270}
\end{center} 
\caption{(Color online) Same as Fig.~\protect\ref{fig4} with FMM}
\label{fig7}
\end{figure} 
\begin{figure}[h]
\begin{center}
\epsfig{file=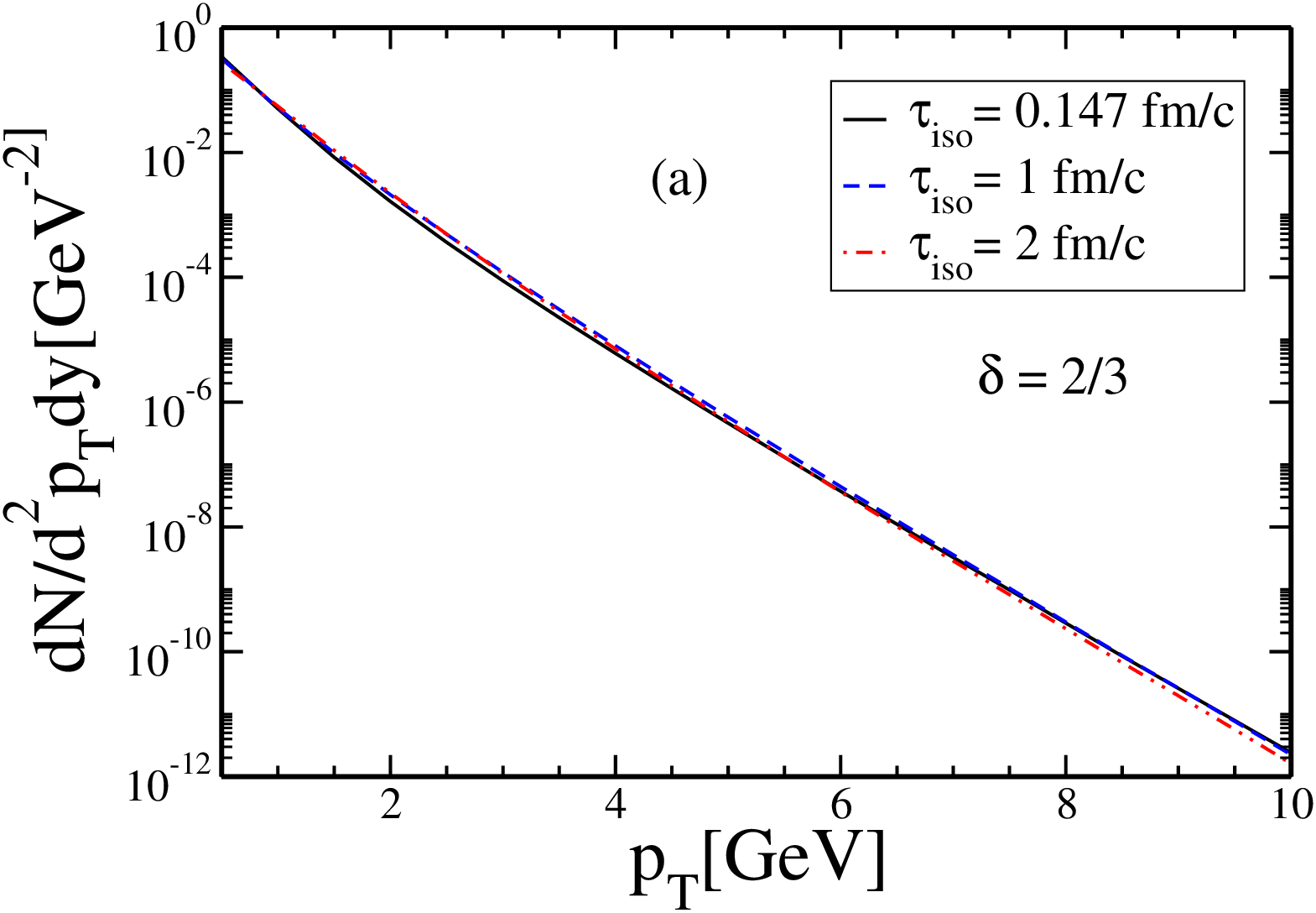,width=7cm,height=8cm,angle=270}
\epsfig{file=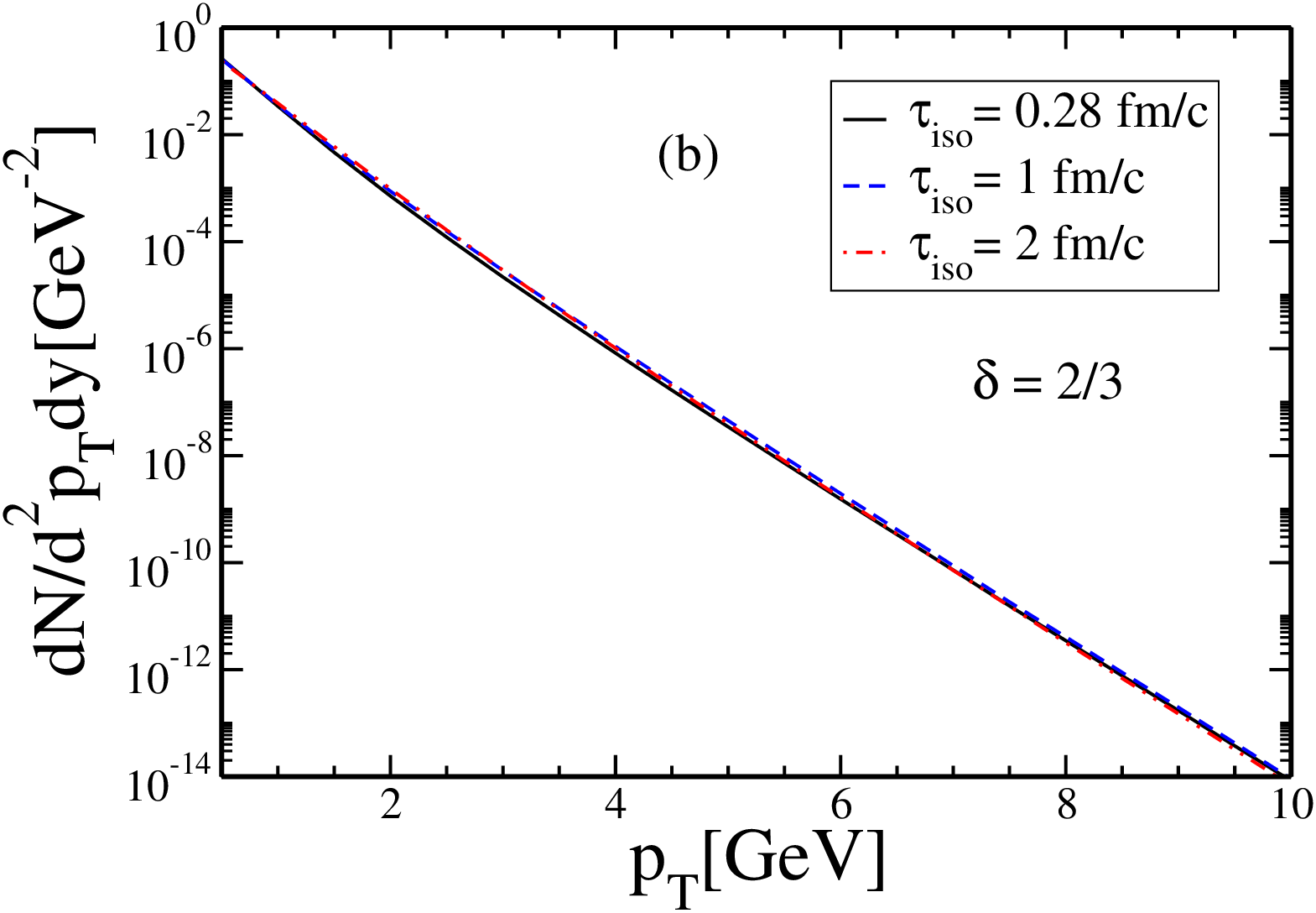,width=7cm,height=8cm,angle=270}
\end{center} 
\caption{(Color online) Same as Fig.~\protect\ref{fig5} with FMM}
\label{fig8}
\end{figure} 
\begin{figure}[h]
\begin{center}
\epsfig{file=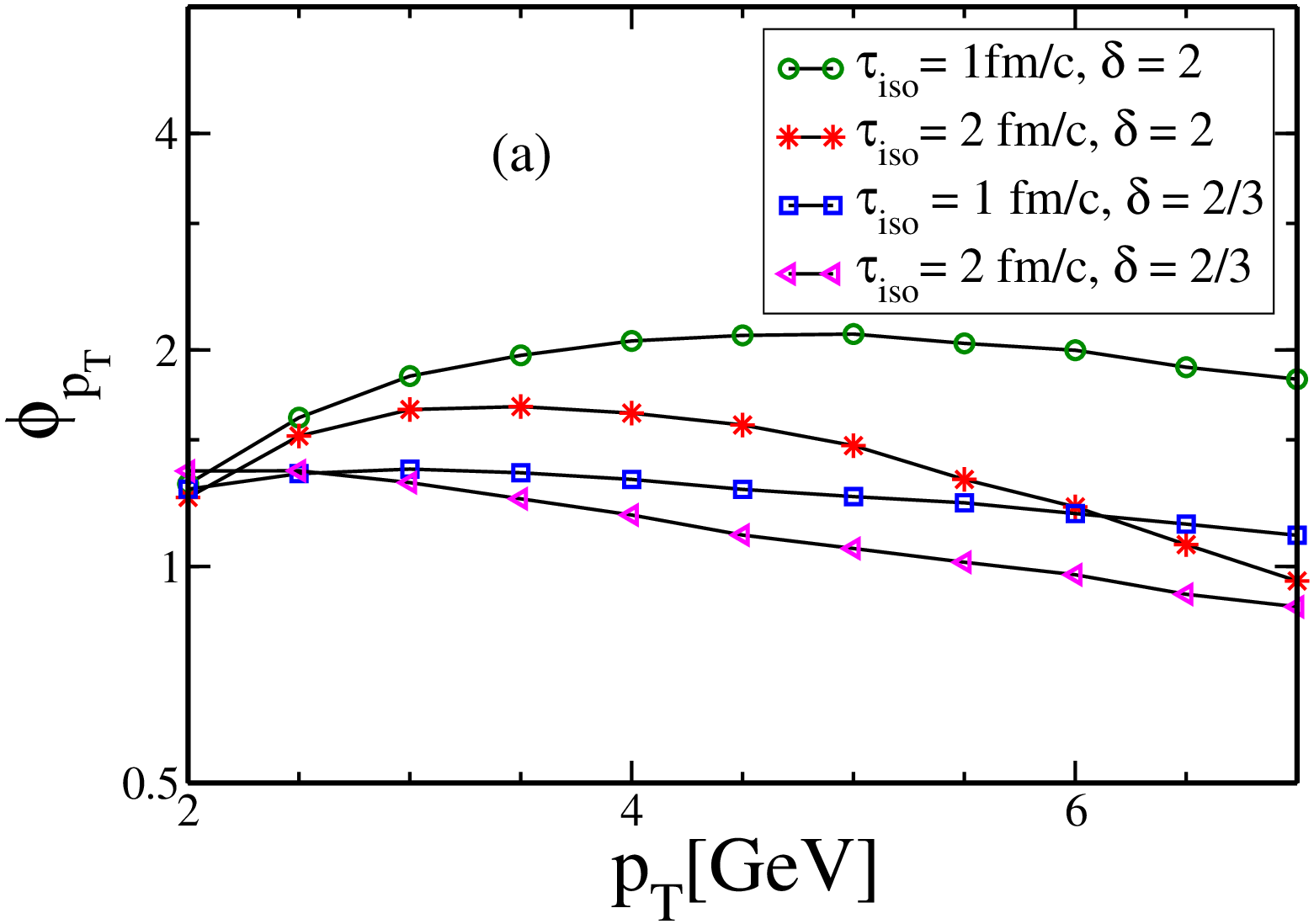,width=7cm,height=8cm,angle=270}
\epsfig{file=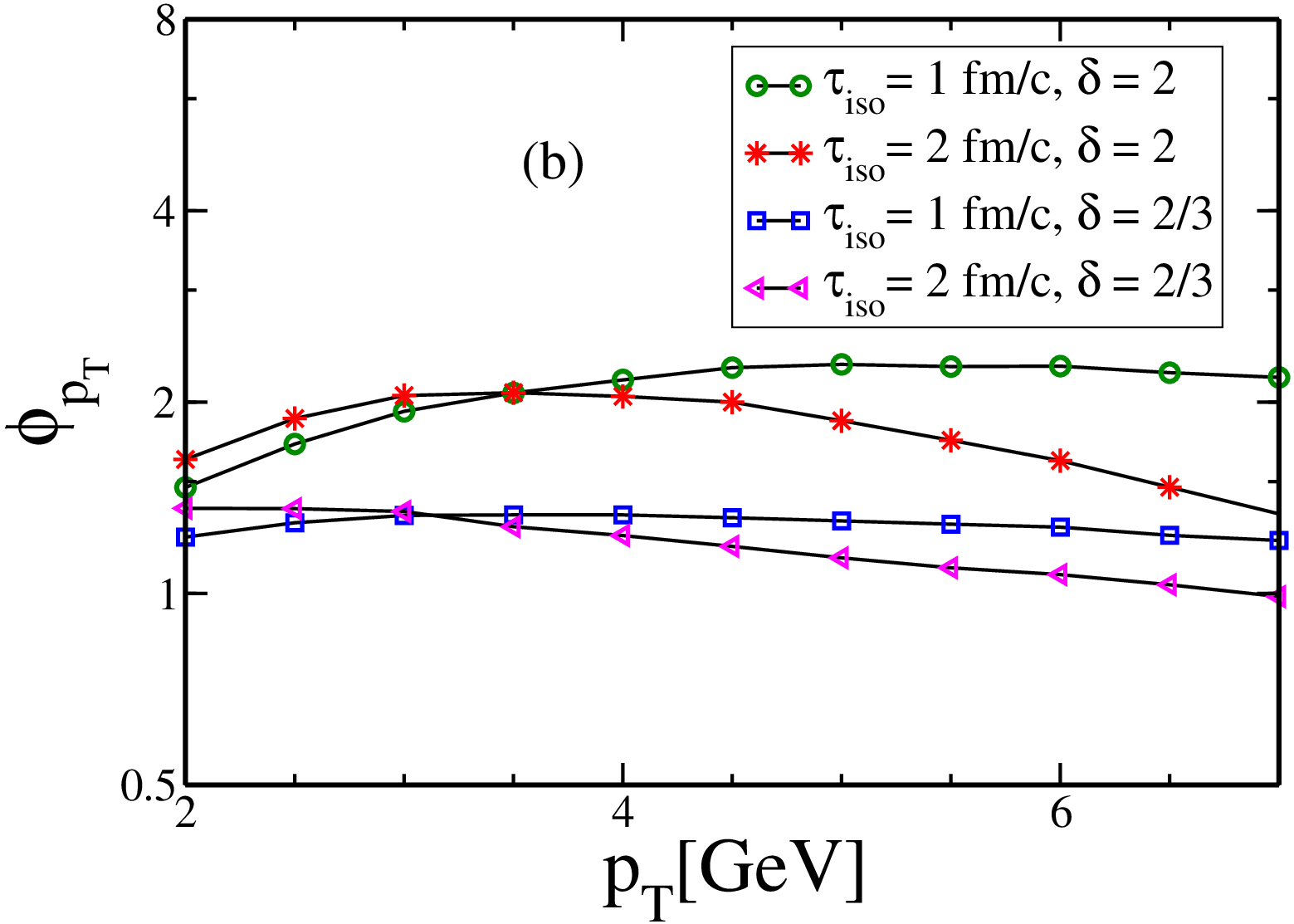,width=7cm,height=8cm,angle=270}
\end{center} 
\caption{(Color online) Same as Fig.~\protect\ref{fig6} with FMM}
\label{fig9}
\end{figure} 

To numerically evaluate the medium photon yield in an anisotropic QGP
we must have the knowledge of space-time dependence of anisotropy
parameter, $\xi(\tau)$, and hard momentum scale, $p_{hard}(\tau)$ as 
discussed earlier. To include the time dependence of $\xi$ and $p_{hard}$, 
we will use Eqs.~({\ref{xirho}}) and ({\ref{utau}}) described in the 
previous section (see Ref.~\cite{mauricio,prl100} for details). 
 
We will now discuss contributions to the total photon yield
due to medium photon spectrum from anisotropic QGP with initial
conditions that might be achieved at 
RHIC and LHC energies. In what follows we shall consider two 
interpolating models: (i) {\em free-streaming interpolating} 
model ($\delta =2 $) and (ii) {\em collisionally-broadened interpolating} 
model ($\delta =2/3$). To cover the uncertainties in the initial 
conditions for a given beam energy, we consider two sets of 
initial conditions, one at a relatively high temperature 
of $T_i=0.446$ GeV and a relatively 
short initial time of $\tau_i=0.147$ fm/c denoted hereafter by Set-I, 
and another at a lower temperature 
$T_i=0.350$ GeV and somewhat later initial time of $\tau_i=0.28$ fm/c denoted
hereafter by Set-II. 
The {\em free-streaming interpolating} model with fixed initial 
conditions (FIC) leads to entropy generation~\cite{mauricio} and, 
therefore, the allowed values of $\tau_{\rm iso}$, in this case, is much 
more constrained. On the other hand, for 
{\em collisionally-broadened interpolation} model, the upper bound 
on $\tau_{\rm iso}$ is much larger due to small amount of entropy 
generation. The estimates of upper bounds on $\tau_{\rm iso}$ for 
both the models are listed in Ref.~\cite{mauricio}.

We also invoke the conditions of fixed final multiplicity (fixed entropy)
to calculate the photon yield from the plasma. In such case, the initial
hard momentum scale ($p_{hard} (\tau_i)$) will be $\tau_{\rm iso}$ 
dependent, i.e. the larger the value
of $\tau_{\rm iso}$ smaller will be the initial hard momentum scale. Therefore,
the enhancement in the yield will be less as compared to the case of
fixed initial conditions. Moreover, as we shall see, for a given condition
the enhancement will be lower in case of {\em collisionally-broadened} 
model as $\delta$ is smaller (close to isotropic expansion).

Fig.~\ref{fig1} shows the medium photon production 
rate (static) as a function of photon energy, $E$, for two different 
values of the anisotropy parameter, $\xi$, and two different values 
of photon angle, $\theta$. We have used $\alpha_{s}=0.3$ and 
$p_{hard}=0.446$ GeV. It is clear from Fig.~\ref{fig1} that 
the photon rate is more suppressed in the direction parallel to 
the direction of anisotropy compared to the transverse direction 
for an anisotropic QGP. This is due to the fact that for $\xi>0$, 
decreasing value of photon angle ($\theta$) implies decreasing 
value of medium parton density and hence decreasing photon rate as can 
be seen from Eq.~(\ref{dist_an}). This is also clear from 
Fig.~\ref{fig2}, which shows the photon rate as a 
function of photon angle, ($\theta$), for $E=2$ 
GeV for two different values of $\xi$. It is observed that for a given
anisotropy parameter, the value of the photon rate is larger in the 
transverse direction. However, the photon rate in an oblate anisotropic 
medium ($\xi>0$) is always small compared to the photon rate 
in an isotropic medium. 
This can be attributed to the fact that an oblate anisotropic 
distribution (Eq. \ref{dist_an}) 
i.e $+ve$ value of $\xi$  corresponds to the suppression of 
medium parton density. As a result, introduction of anisotropy 
(with $+ve$ value of anisotropy parameter ($\xi$)), in general, 
leads to a suppression of medium photon rate (static).

After the very short discussion of the medium photon rate in an
anisotropic QGP, we will now discuss the effects of pre-equilibrium
momentum space anisotropy of the QGP on the thermal photon
spectrum. To illustrate the effects of pre-equilibrium momentum space
anisotropy of QGP on the parton densities, in Fig.~\ref{fig3}, we have
plotted the ratio $(f_{aniso}/f_{iso})$ (for ${\bf |k|} = 5$ GeV) as a
function of the angle between ${\bf k}$ (the momentum of the
plasma parton) and {\bf $\hat n$} (the direction of anisotropy) 
for different values of intermediate time with $\tau_{iso}=0.5$
fm. Here $f_{aniso}({\bf k},~p_{hard})$ is the parton
  distribution function for a QGP with pre-equilibrium momentum space
  anisotropy and $f_{iso}({\bf k},~p_{hard})$ is the parton
  distribution function for a QGP without any pre-equilibrium
  momentum space anisotropy. The anisotropic distribution functions,
  in Fig.~\ref{fig3}, are calculated in the framework of fixed initial
    condition {\em free-streaming interpolation} model with $\delta=2$. 
Fig.~\ref{fig3} clearly shows an enhancement of anisotropic distribution in the
transverse direction and suppression in the longitudinal direction for
$\tau<\tau_{iso}$. The enhancement in the transverse direction is a
consequence of the enhancement of the hard momentum scale, $p_{hard}$, for
$\tau<\tau_{iso}$ . However, in the longitudinal direction, the suppression 
due to the nonzero value of the anisotropy parameter does not stand over 
the enhancement due to the enhanced value of hard momentum scale.

\subsection{Photon spectrum for fixed initial conditions with $\delta =2$ 
and $\delta = 2/3$}

In this section we will present the medium photon spectrum assuming
  fixed initial condition and {\em free streaming} 
({\em collisionally-broadened}) {\em interpolating} model
  with $\delta=2~(2/3)$. In Fig.~\ref{fig4} we have presented the
  medium photon yield in the mid rapidity ($\theta_{\gamma}=\pi/2$) 
as a function of photon transverse momentum at 
  for those two sets of fixed initial
  conditions. Fig.4a (4b) corresponds to $T_i= 0.446~(0.350)$ GeV and
  $\tau_i= 0.147~(0.28)$ fm/c. In estimating these results, we have used
  $\alpha_s=0.3$. Different lines in Fig.~\ref{fig4} correspond to 
different isotropization times, $\tau_{iso}$. We observe enhancement of photon
yield when $\tau_{iso}>\tau_i$. The enhancement of photon yield in 
the transverse directions is due to the fact that momentum-space 
anisotropy enhances the density of plasma partons moving at the mid 
rapidity as can be seen from Fig.~\ref{fig3}. 

In Fig.~\ref{fig5} , we have shown the predicted photon yields 
(at the mid rapidity) for a given beam energy (with two different 
sets of initial conditions) assuming the time dependence of the 
hard momentum scale and anisotropy parameter as given by 
Eq. \ref{xirho} with $\delta=2/3$. It is important to mention 
that $\delta=2/3$ corresponds to {\em collisionally-broadened interpolating} 
model. Fig.~\ref{fig5} shows enhancement of photon yield as 
we increase the isotropization time. However, compared to the FIC 
{\em free-streaming interpolating} model, discussed in the previous para, 
here, the enhancement of the photon yield is small. This is due to the 
fact that in the case of {\em collisionally-broadened interpolating} 
model we have included the possibility of momentum space broadening of 
the plasma partons due to interactions. 
{\em Collisionally broadened interpolating} model is close to thermal
equilibrium. As a result the yield is close to the equilibrium value.

To quantify the effect of isotropization time, we 
define {\em photon enhancement} factor $\Phi_(p_T)$ as \cite{mauricio}:
\begin{equation}
\Phi(\tau_{iso},\theta)=\left(\frac{dN(\tau_{iso})}
{dyd^2p_T}\right)_{\theta}/\left(\frac{dN(\tau_{iso}=\tau_i)}
{dyd^2p_T}\right)_{\theta},
\end{equation}
which is plotted in Fig.~\ref{fig6} as a function of the photon 
transverse momentum.
Fig.~\ref{fig6}a (6b) for $T_i=0.446$ GeV and $\tau_i=0.147$
fm/c ($T_i=0.350$ GeV and $\tau_i=0.28$ fm/c) shows that for the fixed 
initial condition {\em free-streaming interpolating} model, 
the photon yield at $p_T=5$ GeV is enhanced 
by a factor of $13.8~(11.5)$ at mid rapidity at $\tau_{iso}= 2$ fm/c.    
Fig.~\ref{fig6} also includes the photon enhancement factor for FIC 
collisionally broadened interpolating model. However, in the frame work 
of FIC {\em collisionally-broadened interpolation} 
model, enhancement of photon yield is marginal at $\tau_{iso}= 2$ fm/c.

\begin{figure}[h]
\begin{center} 
\epsfig{file=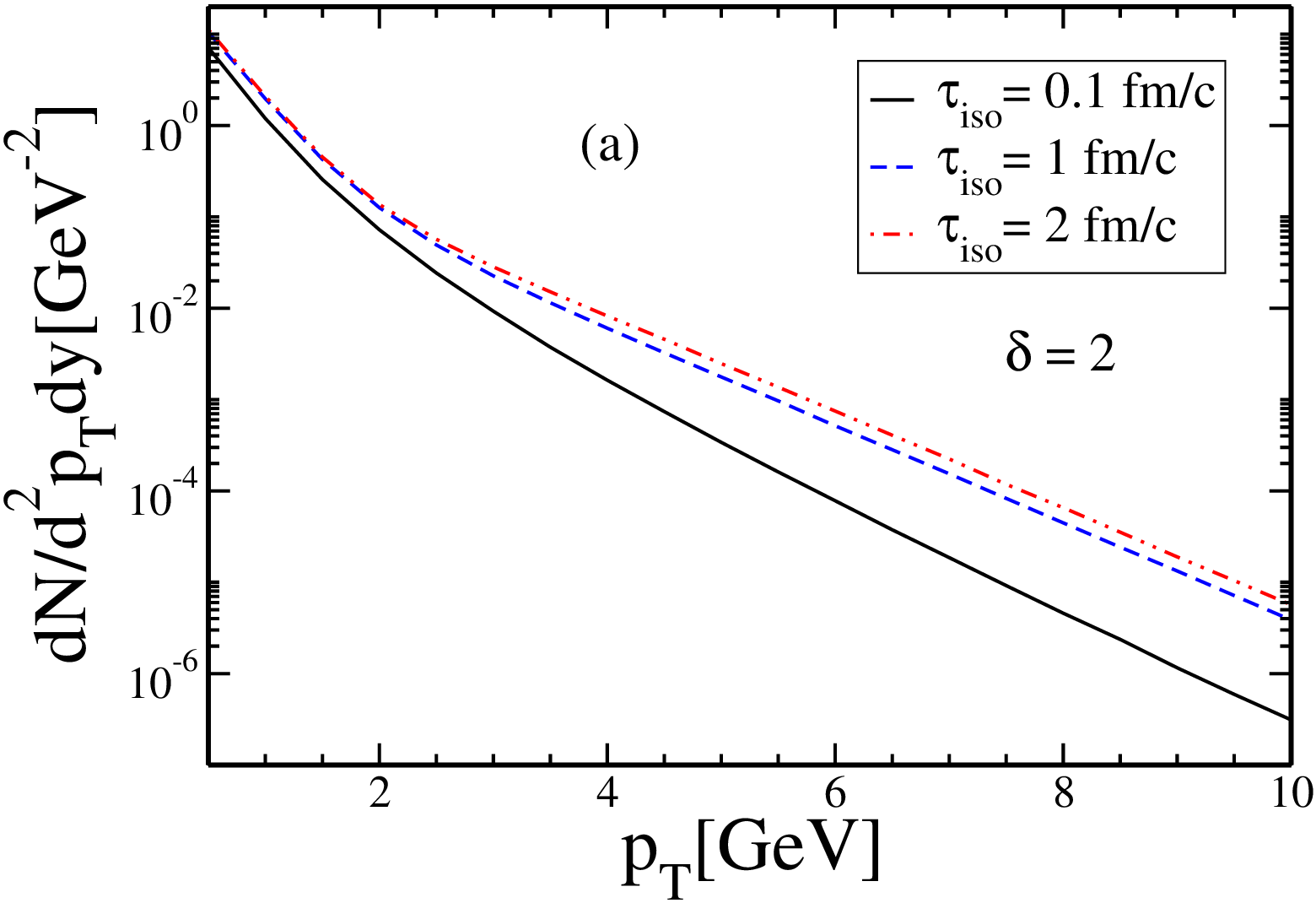,width=7cm,height=8cm,angle=270}
\epsfig{file=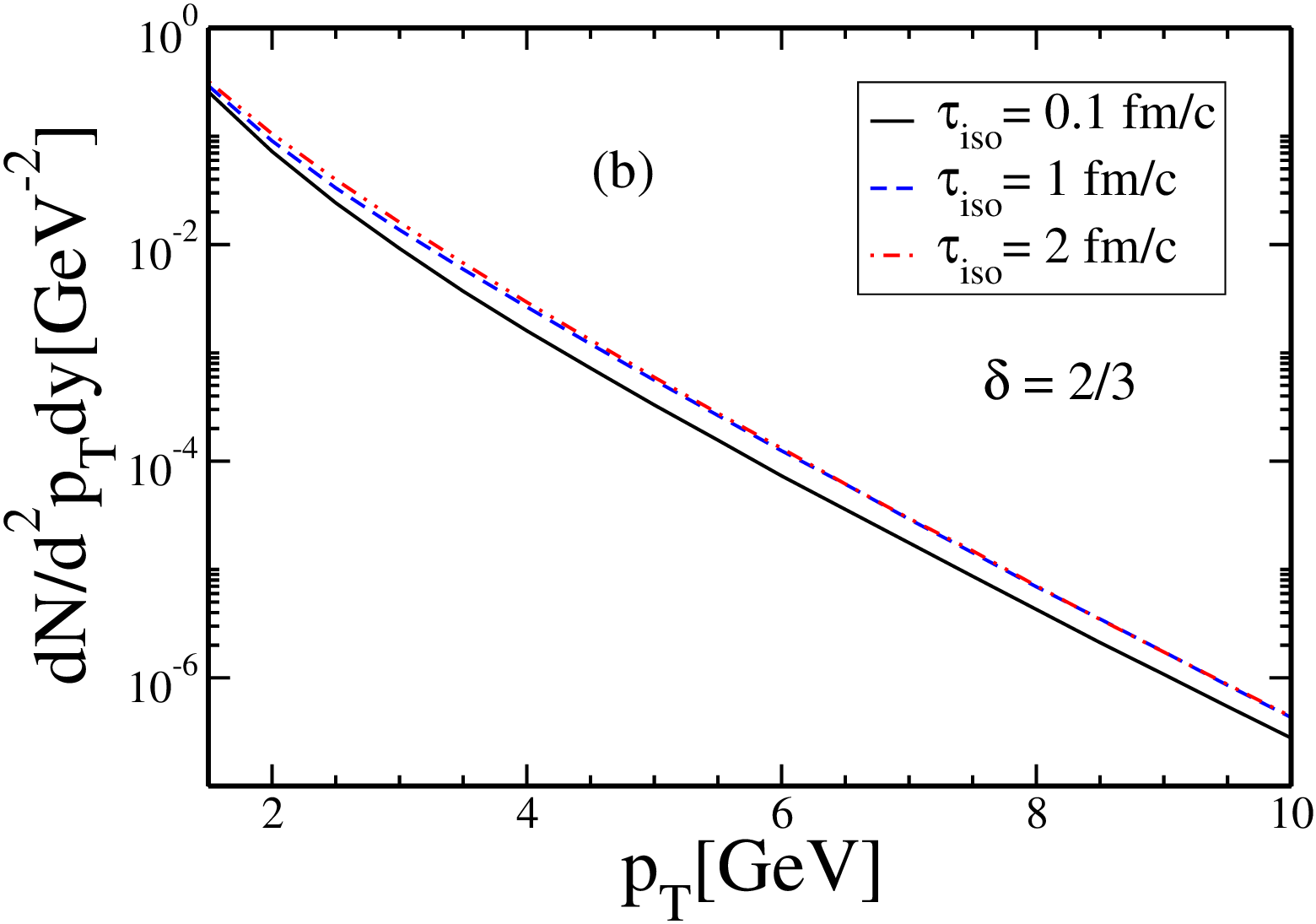,width=7cm,height=8cm,angle=270}
\end{center} 
\caption{(Color online) Medium photon spectrum, $dN/d^2p_Tdy$, at 
$\theta = \pi/2 (y = 0)$.(a) ((b)) corresponds to 
{\em free streaming interpolating model} 
({\em collisionally-broadened interpolating model}) for 
three different values of isotropization 
time, $\tau_{iso}$, with initial temperature $T_i = 0.828$ 
GeV and initial time $\tau_i= 0.1$ fm/c.}
\label{fig10}
\end{figure} 
\begin{figure}[h]
\begin{center}
\epsfig{file=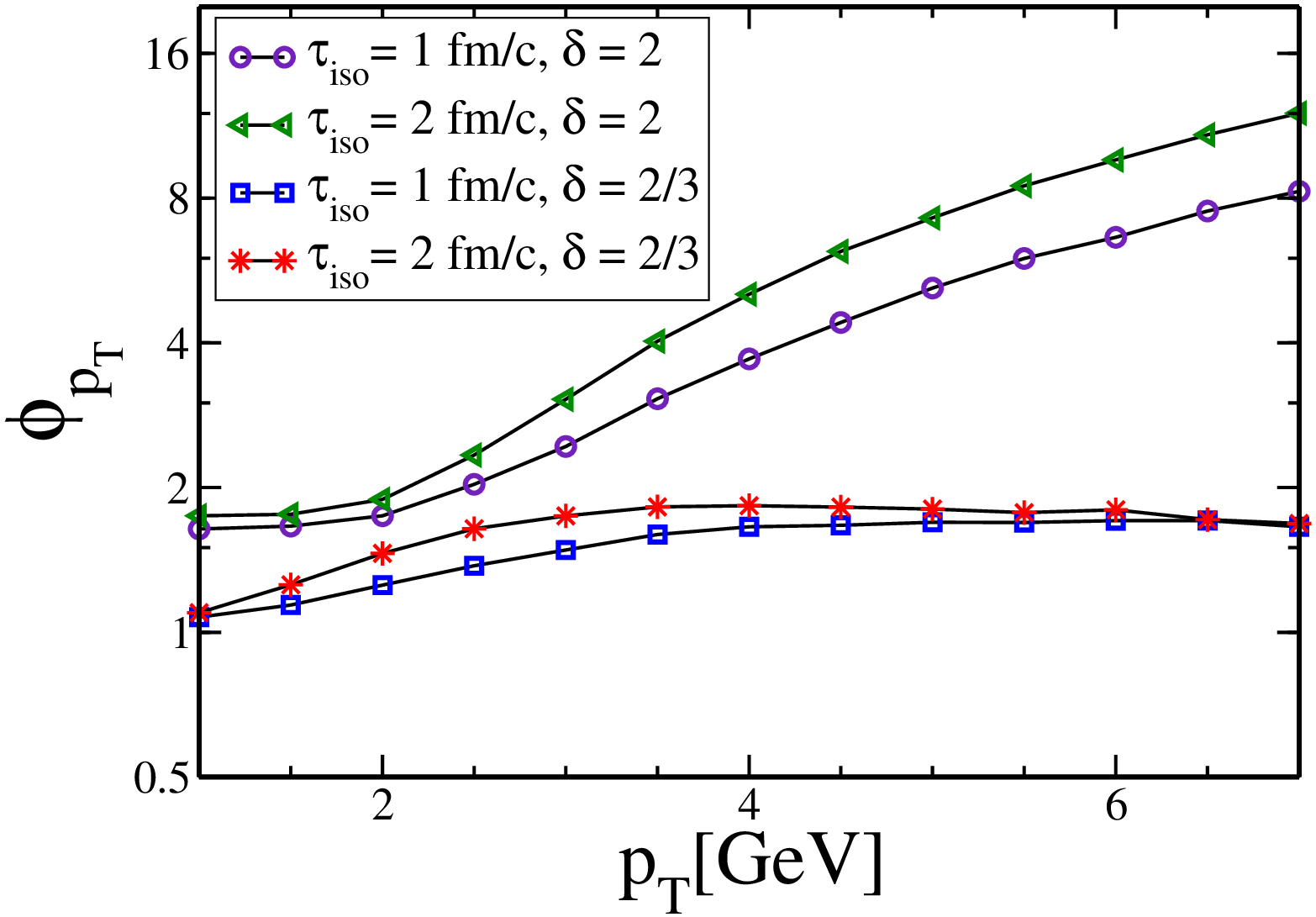,width=7cm,height=8cm,angle=270}
\end{center} 
\caption{(Color online){\em Photon enhancement factor} 
$(\phi_{p_T})$ as a function of 
photon transverse momentum $(p_T)$ for the {\em free streaming} and 
{\em collisionally-broadened interpolating} models ($\delta=2,2/3$) 
with fixed initial condition for two different values of 
isotropization time ($\tau_{iso}$) with $T_i= 0.828$ GeV
and $\tau_i=1$ fm/c.} 
\label{fig11}
\end{figure} 

\subsection{Photon spectrum for fixed final multiplicity (FMM) with $\delta =2$ 
and $\delta = 2/3$}

In the previous section we have used the two interpolating models which
assumes fixed initial conditions. However, enforcing fixed initial
condition results in generation of
particle number and enhance the final multiplicity. Starting with a
fixed initial condition, the interpolating model 
(discussed in section $2.2$) corresponds to a hard momentum 
scale (for $\tau>>\tau_{iso}$) which
is larger by a factor of $[{\cal R}((\tau_{iso}/\tau_i)^{\delta}-1)]^{0.25}$ 
compared to the momentum scale results from the 
free hydrodynamic expansion of a system (with the same initial
condition) from the beginning. To ensure fixed final multiplicity in
this model one has to redefine $\bar{\cal U}(\tau)$ in 
Eq. \ref{xirho} as in ~\cite{mauricio}:
$$ 
\bar{\cal U}(\tau)=\frac{{\cal U}(\tau)}{{\cal U}(\tau_i)}\left[{\cal R}
((\tau_{iso}/\tau_i)^{\delta}-1)\right]^{-3/4}(\tau_{i}/\tau_{iso})
$$     
This redefinition corresponds to a lower initial hard momentum scale 
(${\em p_{hard}(\tau_i)}<T_i$) for $\tau_{iso}>\tau_i$. Larger value
of isotropization time corresponds to lower initial hard momentum scale. 

The transverse photon yields assuming FMM and 
{\em free-streaming interpolating} model are presented 
in Fig.~\ref{fig7} as a function of
  photon transverse momentum for different values of isotropization
  times with the initial conditions Set I (Fig.7a) and
Set II (Fig.7b). In the case of fixed initial condition ({\em free-streaming}
  or {\em collisionally-broadened}) interpolating model we have
  obtained significant enhancement of transverse photon yield due to
  the pre-equilibrium momentum space anisotropy of QGP. However,
  fixing the final multiplicity significantly reduce the effect of
  pre-equilibrium anisotropic phase. 
Moreover as a result of fixing final multiplicity we obtain a suppression of
photon yield in the high $p_T$ region. One more
  interesting consequence of fixing final multiplicity is that larger
  isotropization time corresponds to suppression in the
  high $p_T$ region and less enhancement in the intermediate $p_T$
  region. This is due to the fact that to ensure the fixed final
  multiplicity in the pre-equilibrium {\em free streaming interpolating}
  model, one has to reduce the initial hard momentum scale or equivalently the
  initial energy density. 

The photon yield for two sets of initial conditions, 
assuming {\em collisionally-broadened} $(\delta= 2/3)$ pre-equilibrium 
phase of QGP with fixed final multiplicity is displayed in 
Fig.~{\ref{fig8}}. It is observed that fixing final multiplicity 
significantly reduce the effects of {\em collisionally-broadened} 
pre-equilibrium phase of QGP. As discussed in the previous section, 
this is again due to the fact that in order to maintain fixed final 
multiplicity for $\tau_{iso}>\tau_i$ the initial energy density 
(or equivalently initial hard momentum scale) has to be reduced.


In order to see the difference between FIC and FMM we plot 
$\Phi_{p_T}$ with FMM condition, in Fig.~{\ref{fig9}} both for 
$\delta=2$ and $\delta=2/3$. 
Since the {\em collisionally-broadened interpolating} model is 
always closer to local-isotropic expansion, this 
results in less photon enhancement compared to the 
{\em free-streaming} pre-equilibrium phase (also see Fig.~\ref{fig6}). 
For the initial condition Set-I (Set-II) we predict an 
enhancement of the order of $1.6~(1.53)$ for $p_T=5$ GeV 
as a result of {\em collisionally-broadened} pre-equilibrium 
phase with fixed final multiplicity as we vary $\tau_{iso}$ from 
$\tau_i$ to 2 fm/c, whereas {\em free-streaming} pre-equilibrium 
phase with fixed final multiplicity corresponds to a enhancement 
(for $p_T=5$ GeV) by a factor of $2.1~(1.13)$. 

\subsection{Photon yield for fixed initial conditions with higher
initial temperature}

To demonstrate the effect of higher initial temperature (might be relevant
for LHC energies) we proceed to calculate the photon yield with FIC for
both the {\em free streaming} ($\delta=2$) and {\em collisionally broadened}
($\delta=2/3$) {\em interpolating} models. The transverse momentum 
distribution of medium photons in central 
rapidity region with $T_i=0.828$ GeV
and $\tau_i=0.1$ fm/c is displayed in Fig.~{\ref{fig10}}.  
Here also {\em free-streaming} pre-equilibrium phase corresponds to more 
enhancement compared to the {\em collisionally-broadened} 
pre-equilibrium phase. In Fig.~\ref{fig11}, we have plotted the photon 
{\em enhancement} factors (for two different values of isotropization 
time) as a function of photon transverse momentum both for the 
{\em free-streaming} and {\em collisionally-broadened} interpolating 
model with fixed initial condition. Fig.~\ref{fig11} shows an enhancement (for
$p_T=5$ GeV) of transverse photon yield by a factor of 
$7.3~(1.8)$ at $\tau_{iso}= 2$ fm/c for fixed initial 
condition {\em free-streaming} ({\em collisionally-broadened}) 
{\em interpolating} model.      

It should be noted that in order to apply our calculations to explain the
experimental data we should add to this the contributions from various
other sources of photon productions. The model used for this purpose
should properly include the effects of tranverse flow as it
affects the photon spectra significantly.
Applications of the present work to calculate the photon
$p_T$ distribution at RHIC and LHC energies incorporating tranverse 
flow effects is in progress and will be reported elsewhere~\cite{LB}.

\section{Conclusion} 

To summarize, we have investigated the effects of the pre-equilibrium
momentum space anisotropy of the QGP on the medium photon
production due to Compton and annihilation processes with various
initial conditions. 
To describe the transition of the plasma from 
initial non-equilibrium state to an isotropic thermalized 
state, we have used two phenomenological models for the time
dependence of the hard momentum scale ($p_{hard}$) and plasma 
anisotropy parameter ($\xi$). The first model is based on
the assumption of fixed initial condition. However, enforcing fixed
initial condition causes entropy generation. Therefore,
we have also considered a second model which assumes fixed final
multiplicity. Both the possibilities of {\em free streaming} and 
{\em collisionally broadened} pre-equilibrium phase
of the QGP are considered. To cover
the uncertainties in the initial conditions for 
for a given beam energy, we have used two sets of initial conditions, one at
a relatively high temperature and a relatively short initial time and
other at a lower temperature and somewhat later initial time.

We estimate the $p_T$ distribution of photons for different
isotropization times in the frame work of these phenomenological
models. It is found that, for fixed initial condition, a {\em
  free-streaming} interpolating model can enhance the thermal photon
yield by an order of magnitude at higher $p_T$. However, 
for {\em collisionally broadened} pre-equilibrium phase with fixed 
initial condition, the enhancement of photon yield is not that much. 
This is due to the fact that the energy density, hard 
momentum scale and anisotropy parameter corresponding to the 
{\em collisionally broadened} interpolating model are always 
closer to the local isotropic expansion. Since fixing the final
multiplicity reduce the initial hard momentum scale or equivalently the
initial energy density, the enhancement of the photon yield is
significantly reduced (both for the {\em free streaming} and {\em
  collisionally broadened} interpolating models). 
Moreover, for fixed final multiplicity, we predict
a suppression of thermal photon yield in the low and high transverse
momentum region as has been observed in the case of dilepton invariant mass
spectrum for anisotropic plasma~\cite{mauricio}. 

\end{document}